\documentclass{article}
\usepackage[utf8]{inputenc}

\usepackage[singlelinecheck=false]{subcaption}
\usepackage[margin=1in]{geometry}
\usepackage[utf8]{inputenc}
\usepackage{bbm, bm}
\usepackage{graphicx}
\usepackage{color, xcolor}
\usepackage{amsmath, amsfonts, amssymb, amsthm, thmtools, mathtools}
\usepackage{algorithm, algpseudocode}
\usepackage[most]{tcolorbox}

\PassOptionsToPackage{hyphens}{url}
\usepackage[colorlinks=true, allcolors=blue]{hyperref}
\usepackage[capitalise,nameinlink]{cleveref}

\usepackage[style=alphabetic, backend=biber, minalphanames=3, maxalphanames=4, maxbibnames=99, maxcitenames=99]{biblatex}

\crefformat{section}{#2\S#1#3}
\crefformat{subsection}{#2\S#1#3}
\crefformat{subsubsection}{#2\S#1#3}
\Crefformat{section}{#2\S#1#3}
\Crefformat{subsection}{#2\S#1#3}
\Crefformat{subsubsection}{#2\S#1#3}

\Crefrangeformat{section}{#3\S\S#1–#2#4}
\Crefmultiformat{section}{#2\S\S#1#3}{ and #2#1#3}{, #2#1#3}{, and #2#1#3}
\declaretheoremstyle[bodyfont=\it,qed=\qedsymbol]{noproofstyle}

\numberwithin{equation}{section}

\declaretheorem[name=Observation,numbered=no]{observation*}

\declaretheorem[numberlike=equation]{theorem}

\declaretheorem[name=Theorem,numbered=no]{theorem*}

\declaretheorem[name=Lemma,numbered=no]{lemma*}

\declaretheorem[name=Corollary,numbered=no]{corollary*}

\declaretheorem[name=Proposition,numbered=no]{proposition*}

\declaretheorem[name=Claim,numbered=no]{claim*}

\declaretheorem[name=Question,numbered=no]{question*}

\declaretheoremstyle[bodyfont=\it]{defstyle}

\declaretheorem[unnumbered,name=Definition,style=defstyle]{definition*}

\declaretheorem[unnumbered,name=Example,style=defstyle]{example*}

\declaretheorem[unnumbered,name=Notation=defstyle]{notation*}

\declaretheorem[unnumbered,name=Construction,style=defstyle]{construction*}

\declaretheoremstyle[
  headfont=\normalfont\bfseries,
  bodyfont=\normalfont,
  postheadspace=1em,
  qed=$\lozenge$
]{rmkstyle}

\declaretheorem[
  style=rmkstyle,
  numberlike=equation
]{remark}

\tcbset{
  mytheorem/.style={
    enhanced,
    breakable,
    colback=blue!5,
    colframe=blue!75!black,
    coltitle=white,
    fonttitle=\bfseries,
    boxed title style={colback=blue!75!black},
    attach boxed title to top left={xshift=2mm,yshift=-2mm},
    boxrule=0.8pt,
    arc=3pt,
    before skip=10pt,
    after skip=10pt,
    top=10pt
  }
}

\declaretheoremstyle[
  headformat={},
  headpunct={},
  bodyfont=\itshape
]{boxedconjstyle}

\declaretheorem[style=boxedconjstyle,name=Conjecture]{conjecture}

\renewenvironment{conjecture}[1][]{%
  \refstepcounter{conjecture}%
  \begin{tcolorbox}[mytheorem,
    title={Conjecture~\theconjecture\if\relax\detokenize{#1}\relax\else\ (#1)\fi}]%
  \itshape
}{%
  \end{tcolorbox}
}

\newcommand{\Cut}{\textsc{Cut}}
\newcommand{\DiCut}{\textsc{Dicut}}
\newcommand{\TwoAnd}{2\textsc{AND}}
\newcommand{\kAnd}{k\textsc{AND}}
\newcommand{\MaxkAnd}{\textsc{Max-}k\textsc{And}}

\newcommand{\MaxThreeAnd}{\textsc{Max-}3\textsc{And}}
\newcommand{\MaxCut}{\textsc{Max-Cut}}
\newcommand{\MaxDiCut}{\textsc{Max-DiCut}}
\newcommand{\MaxCSP}{\textsc{Max-CSP}(\Pi)}

\newcommand{\MaxkXor}{\textsc{Max-}k\textsc{Xor}}

\newcommand{\MaxkSat}{\textsc{Max-}k\textsc{Sat}}

\newcommand{\MaxqCol}{\textsc{Max-}q\textsc{Coloring}}

\newcommand{\MaxqUG}{\textsc{Max-}q\textsc{UniqueGames}}

\newcommand{\val}[2]{\mathsf{val}_{#1}(#2)}
\newcommand{\maxval}[1]{\mathsf{max}\text{-}\mathsf{val}(#1)}

\newcommand{\atriv}{\alpha_{\mathrm{triv}}}

\newcommand{\bcalG}{\boldsymbol{\mathcal{G}}}
\newcommand{\bPhi}{\boldsymbol{\Phi}}

\newcommand{\Not}{\textsc{Not}}

\newcommand{\Alg}{\mathtt{Alg}}
\newcommand{\Output}{\mathtt{Output}}
\newcommand{\Compress}{\mathtt{Compress}}
\newcommand{\Compose}{\mathtt{Compose}}

\newcommand{\DYes}{\mathcal{D}_{\mathrm{Yes}}}
\newcommand{\DNo}{\mathcal{D}_{\mathrm{No}}}

\newcommand{\MaxkMon}{\textsc{Max-}k\textsc{Monarchy}}
\newcommand{\kMon}{k\textsc{Monarchy}}

\newcommand{\MAS}{\textsc{Max-Acyclic-Subgraph}}
\newcommand{\MBtwn}{\textsc{Max-Betweenness}}
\usepackage{singer-macros}
\usepackage{tikz}
\usetikzlibrary{arrows.meta}

\title{Nine lower bound conjectures on streaming approximation algorithms for CSPs}

\author{Noah G.
Singer\thanks{Department of Computer Science, Carnegie Mellon University, Pittsburgh, PA, USA.
Email: \texttt{ngsinger@cs.cmu.edu}.}}
\date{\today}

\begin{document}

\maketitle

\begin{abstract}
    In this column, we overview recent progress by many authors on understanding the approximability of constraint satisfaction problems (CSPs) in low-space streaming models.
    Inspired by this recent progress, we collate nine conjectural lower bounds against streaming algorithms for CSPs, some of which appear here for the first time.
\end{abstract}

\section{Introduction}

Inspired by an open question at the 2011 Bertinoro workshop \cite{IMNO11}, the last decade has seen an explosion of interest in using \emph{streaming algorithms} for \emph{approximating constraint satisfaction problems (CSPs)}.
Some results we know in this area include:
\begin{itemize}
    \item Single-pass lower bounds for $\MaxCut$~\cite{KK15,KKS15,KKSV17,KK19},
    \item Multi-pass lower bounds for $\MaxCut$~\cite{AKSY20,AN21,CKP+23,FMW25} and other CSPs~\cite{FMW25-dichotomy},
    \item Algorithms and lower bounds for approximating $\MaxDiCut$~\cite{GVV17,CGV20,SSSV23-random-ordering,SSSV23-dicut,SSSV25},
    \item Quantum algorithms and lower bounds for $\MaxCut$ and $\MaxDiCut$~\cite{KP22,KPV24,KPV25},
    \item Results on other specific CSPs, including unique games (\cite{GT19}), monarchy-like predicates (\cite{CGS+22-monarchy}), and $\MaxkAnd$ (\cite{Sin23-kand}),
    \item Dichotomy theorems and results for general CSPs~\cite{CGSV21-boolean,CGS+22-linear-space,CGSV24},
\end{itemize}
and various other results, including lower bounds for \emph{ordering CSPs} (including $\MAS$ and $\MBtwn$)~\cite{SSV24-jour-version},
for solving CSPs \emph{exactly}~\cite{Zel11,SW15,KPSY23},
and for solving CSPs approximately on \emph{dense} instances~\cite{BDV18}.
See the surveys \cite{Sin22,Sud22,Vel23} for some (perhaps already out of date!) exposition.

In this column, I highlight nine ``frontier'' conjectures that have emerged in recent works in this area
(and give some brief overviews of the notions needed to understand the questions).
I will do my best to cite conjectures if they already appear in published work; some appear may here for the first time.

\paragraph*{Acknowledgements.}
These conjectures have emerged out of discussions and papers with several wonderful collaborators of mine including Raghuvansh Saxena, Madhu Sudan, and Santhoshini Velusamy.
I would also like to thank Matthew Ding, William He, and Michael Kapralov for useful discussions;
Bill Gasarch for his helpful proofreading and feedback on this column;
and my advisor Ryan O'Donnell for his support.
During the time of writing, I was supported by an NSF Graduate Research Fellowship (Award DGE2140739).

\section{Constraint satisfaction problems}

\emph{Constraint satisfaction problems (CSPs)} capture a broad class of computational problems. 
In this column, we will only consider \emph{maximization} CSPs;
these include numerous well-studied problems such as $\MaxCut$, $\MaxDiCut$, $\MaxkSat$, $\MaxkXor$, $\MaxqCol$, and $\MaxqUG$.\footnote{
    By ``maximization'', we mean that the goal is to determine (or approximate) the maximum satisfiable fraction of constraints.
    A related problem is deciding whether there exists an assignment satisfying all constraints;
    a dichotomy theorem for such problems was shown in the seminal work of~\textcite{Sch78},
    who showed that every (Boolean) such problem is either in $\P$ or is $\NP$-complete.
    \textcite{Cre95} established a similar theorem for Boolean maximization CSPs.}
These problems, and their hardness of approximation, have been studied extensively throughout complexity theory;
see e.g.~\cite{JS87,Cre95,GW95,TSSW00,Has01,BHPZ23} (a small, chronological sampling of many, many papers).
Maximization CSPs are also intimately connected with the unique games conjecture~\cite{Kho02,KKMO07,Rag08} and with probabilistically checkable proofs~\cite{Din07}.

In this column, we restrict further to the case of \emph{Boolean} CSPs,
which keeps things interesting while simplifying notation.
Here is the general setup we consider.
Let $k \in \N$ be a (typically small) number, the \emph{arity},
and let $\Pi \subseteq (\{0,1\}^k)^{\{0,1\}}$ denote a set of \emph{predicate} functions $\{0,1\}^k \to \{0,1\}$.
For $n \in \N$, a \emph{constraint} is a tuple $C = (j_1,\ldots,j_k;\pi)$ for distinct $j_1,\ldots,j_k \in [n]$ and $\pi \in \Pi$.
An \emph{assignment} is a vector $x = (x_1,\ldots,x_n) \in \{0,1\}^n$, and $x$ \emph{satisfies} the constraint $C = (j_1,\ldots,j_k;\pi)$ iff $\pi(x_{j_1},\ldots,x_{j_k}) = 1$.
An \emph{instance} $\Phi$ consists of a list of constraints, and the \emph{value} of an assignment $x \in \{0,1\}^n$ on $\Phi$ is \[
\val{\Phi}{x} \coloneqq \Pr_{\bC \sim \Phi} \bracks*{ x\text{ satisfies } \bC },
\]
(here the distribution on $\bC$ is uniform over all constraints, or sometimes $\Phi$ might also specify a weight distribution).
The goal of the problem $\MaxCSP$ is to approximate the quantity \[
\maxval{\Phi} \coloneqq \max_{x \in \{0,1\}^n} \val{\Phi}{x}, \]
the maximum value of any assignment.
Specifically, we say $v \in [0,1]$ is an \emph{$\alpha$-approximation} for $\MaxCSP$ if $\alpha \cdot \maxval{\Phi} \le v \le \maxval{\Phi}$.
We let \[
\atriv(\Pi) \coloneqq \lim_{n \to \infty} \bracks*{ \inf_{\Phi \text{, $\MaxCSP$ inst. on }n\text{ vars.}} \maxval{\Phi} }
\]
denote the so-called ``trivial approximation ratio'' for $\Pi$;
this is, informally, the best possible lower bound on $\maxval{\Phi}$ which does not actually depend on $\Phi$.
Note that for every $\epsilon > 0$ and large enough $n$, the value $\atriv(\Pi)-\epsilon$ is always a $(\atriv(\Pi) - \epsilon)$-approximation for $\MaxCSP$.
The complexity-theoretic question we are interested in is: Are $(\atriv(\Pi)+\epsilon)$-approximations possible, and if so, how large can $\epsilon$ be?

\paragraph{Examples of CSPs.}
The definition in the previous paragraph captures a wide array of CSPs, but it turns out that even very simple special cases are quite interesting from a complexity-theoretic perspective.
The simplest interesting CSP is $\MaxCut$, wherein $k = 2$ and $\Pi = \{\Cut\}$ where $\Cut(x_1,x_2) \coloneqq x_1 \oplus x_2$ (where $\oplus$ is the binary \textsc{Xor} operation).
(Equivalently, $\Cut(x_1,x_2) = 1$ iff $x_1 \ne x_2$.)
The second simplest CSP is $\MaxDiCut$, where again $k=2$ but $\Pi = \{\DiCut\}$ where $\DiCut(x_1,x_2) \coloneqq 1[x_1 = 1 \wedge x_2 = 0]$
(equiv., $\DiCut(x_1,x_2) = x_1 \wedge \overline{x_2}$).

Here is another interesting example:
For $k \in \N$ and $b \in \{0,1\}^k$, let $\Not_b : \{0,1\}^k \to \{0,1\}^k$ be the function $\Not_b(x_1,\ldots,x_k) \coloneqq (x_1 \oplus b_1,\ldots,x_k \oplus x_k)$.
(It is useful to think of $\Not_b$ as placing negations on some variables.
For instance, $\Not_{011}(x_1,x_2,x_3) = (x_1,\overline{x_2},\overline{x_3})$.)
Let $\kAnd : \{0,1\}^k \to \{0,1\}$ be the function $\kAnd(x_1,\ldots,x_k) = \bigwedge_{i=1}^k x_i$.
In the $\MaxkAnd$ problem, $\Pi = \{\kAnd \circ \Not_b : b \in \{0,1\}^k\}$.
(For instance, $\TwoAnd \circ \Not_{01} = \DiCut$.)

Note that $\MaxCut$ and $\MaxDiCut$ both involve only one predicate.
Further, the predicate $\Cut$ is symmetric to reordering its inputs.
Thus, it simplifies notation to imagine $\MaxCut$ constraints as unordered pairs $\{j_1,j_2\}$ and $\MaxDiCut$ constraints as ordered pairs $(j_1,j_2)$.
Correspondingly, we can view the input to a $\MaxCut$ problem as an undirected graph $\calG$ on vertex-set $[n]$
and the input to a $\MaxDiCut$ problem as a directed graph $\calG$ on $[n]$, and refer to the constraints in these problems as edges.

It is not hard to check that $\atriv(\MaxCut) = \frac12$, $\atriv(\MaxDiCut) = \frac14$, and $\atriv(\MaxkAnd) = \frac1{2^k}$.
(For instance, for $\MaxCut$, a random graph $\bcalG$ with $\Omega_\epsilon(n)$ edges typically has $\maxval{\bcalG} \le \frac12+\frac{\epsilon}2$,
while for every graph $\calG$, a uniformly random assignment $\bx \in \{0,1\}^n$ has $\Exp \val{\calG}{\bx} = \frac12$.)
\textcite{GW95} famously showed that very nontrivial ($\frac{2}{\pi} \max_{0\le \theta \le \pi} \frac{\theta}{1-\cos \theta} \approx 0.878$-)approximations to $\MaxCut$ are possible in polynomial time,
and subsequent decades have seen extensive work on the polynomial-time approximability of these problems;
beating this ratio is known to be $\NP$-hard assuming the unique games conjecture~\cite{KKMO07}.

\section{Streaming algorithms}

In this column, we are interested in the $\MaxCSP$ problem in a specific algorithmic model, namely, the \emph{streaming} model.
In this model, the algorithm has the following kind of access to an input instance $\Phi$: 
First, it receives the number of variables $n$ in $\Phi$,
and then it receives the constraints $C_1,\ldots,C_m$ in $\Phi$ one by one (in a possibly adversarial order).
Between receiving constraints $C_i$ and $C_{i+1}$, the algorithm may only store $s$ bits of internal memory state, where $s$ is a (typically small) function of $n$.\footnote{
    In this column, space is always measured in bits.}
At the end of the stream, the algorithm is asked to output an $\alpha$-approximation to $\maxval{\Phi}$;
the complexity-theoretic question is how much space is required to achieve particular values of $\alpha$.

Formalizing this is not difficult: For fixed $n$, a deterministic algorithm for $\MaxCSP$ is a pair $(\Alg : \calC \times \{0,1\}^s \to \{0,1\}^s, \Output : \{0,1\}^s \to [0,1])$ where $\calC$ is the set of possible constraints for $\Pi$.
The algorithm starts at some initial state $S_0$; as constraints arrive, the state $S_{j+1} \gets \Alg(C_j, S_j)$ updates iteratively, and the final output is $\Output(S_j)$.
A randomized algorithm for $\MaxCSP$ is a distribution over deterministic algorithms.
In this column, we are concerned with algorithms achieving, say, $\frac23$ probability of outputting correct approximations.

Standard sparsification arguments show that for any $\MaxCSP$ instance $\Phi$,
if $\bPhi$ is a ``subsampled'' random instance with $m = \Theta(n/\epsilon^2)$ constraints, each of which is sampled i.i.d. uniformly from $\Phi$,
then w.h.p. $|\maxval{\bPhi} - \maxval{\Phi}| \le \epsilon$.
This essentially gives the following algorithmic result:

\begin{theorem}[Folklore, see e.g.~\cite{CGS+22-linear-space}]\label{thm:sparsifier}
    For every constraint family $\Pi$ and $\epsilon > 0$, there is an $(1-\epsilon)$-approximation streaming algorithm for $\MaxCSP$ in $O(n \log n/\epsilon^2)$ bits of space.
\end{theorem}

\begin{remark}
    Our definition of the streaming model makes no assumptions about the algorithm's running time,
    meaning that an algorithm can calculate $\maxval{\bPhi}$ \emph{exactly} (even though this problem is $\mathsf{NP}$-hard).
\end{remark}

On the other end of the spectrum, simply outputting the trivial approximation $\atriv(\MaxCSP)$ uses zero space and achieves an $(\atriv(\MaxCSP)-\epsilon)$-approximation for $\epsilon > 0$.
The ``nontrivial'' regime, therefore, is using space $\omega(1)$ and $o(n \log n)$ to get approximation ratios $\atriv(\MaxCSP) < \alpha \le 1$.
For some CSPs, like $\MaxCut$, it appears that this is essentially impossible~\cite{KKS15,KK19,FMW25},
and the interesting questions are about proving lower bounds with optimal parameters (see \Cref{sec:single-pass max-cut,sec:multi-pass max-cut}).
For other CSPs, like $\MaxDiCut$, nontrivial approximations are possible~\cite{GVV17,CGV20,CGSV24,SSSV23-dicut},
and there are many open questions on tradeoffs between streaming parameters and the approximation ratio (see \Cref{sec:dicut} below).

\paragraph{Variant models.}
There are a few interesting variations on the streaming model we described above.
At times, we make the model more generous to algorithms:
\begin{itemize}
    \item assuming the provided list of constraints in $\Pi$ is uniformly randomly ordered (as opposed to adversarially ordered),
    \item assuming the instance $\Pi$ is ``bounded-degree'', meaning that every individual variable $i \in [n]$ appears in only $O(1)$ constraints,
    \item allowing the algorithm to make multiple passes over the list of constraints $C_1,\ldots,C_m$.
\end{itemize}
Conversely, when proving lower bounds, we might need to make more stringent assumptions on possible algorithms.
Specifically, a \emph{sketching} algorithm is a special type of streaming algorithm describable by functions $\Compress : \calC \to \{0,1\}^s$ and $\Compose : \{0,1\}^s \times \{0,1\}^s \to \{0,1\}^s$ such that $\Alg(S,C) = \Compose(S, \Compress(C))$ and:
\begin{multline*}
\Compose(\Compress(C_1),\Compose(\Compress(C_2),\Compress(C_3))) \\
= \Compose(\Compose(\Compress(C_1),\Compress(C_2)),\Compress(C_3)).
\end{multline*}
Informally, this rules out streaming algorithms that treat constraints differently depending on where they appear in the stream.\footnote{
    The reasons we consider sketching algorithms are twofold.
    Firstly, many natural algorithms for streaming CSPs are sketching algorithms~\cite{GVV17,CGV20,CGSV24,SSSV23-dicut}.
    Secondly, sketching algorithms can be simulated in the \emph{simultaneous communication model}.
    In turn, this model can be simulated by the \emph{sequential communication model} (which can also simulate general streaming algorithms).
    It is often easier to prove lower bounds in the simultaneous model.}

\paragraph{Why streaming CSPs?}
There are a few reasons for why it is so interesting to study the approximability of CSPs via streaming algorithms.
By ignoring time complexity, we can prove (unconditional!) lower bounds against streaming algorithms;
these can be viewed as information-theoretic limits on the extent to which a $\MaxCSP$ instance $\Phi$ can be \emph{compressed} while maintaining enough information to recover $\maxval{\Phi}$.
Indeed, all existing streaming lower bounds we cite in this column are unconditional
and proven via techniques from communication complexity.
At the same time, streaming algorithms can achieve nontrivial approximations for many problems, including $\MaxDiCut$ (\cite{GVV17}).
Progress on streaming algorithms for CSPs has employed ideas from sketching, sampling, local, and distributed algorithms;
in turn, this progress has led to simpler \emph{polynomial-time} approximation algorithms for some problems~\cite{BHP+22}.
See \Cref{rmk:dicut vs cut} below for some (very rough) intuition on why some CSPs admit algorithms in the streaming setting and others do not.

\section{Single-pass, linear(ish)-space streaming lower bounds}\label{sec:single-pass max-cut}

Recall that $\MaxCut$ is the ``simplest interesting'' example of a CSP, and that the trivial approximation threshold for $\MaxCut$ is $\atriv(\Cut) = \frac12$.
It turns out that in the (sublinear-space) streaming setting, doing any better than a trivial $\frac12$-approximation for $\MaxCut$ is very hard.
After a significant line of work~\cite{KK15,KKS15,KKSV17,KK19}, the strongest single-pass lower bounds for $\MaxCut$ which we currently know are the following:

\begin{theorem}[\textcite{KK19}]\label{thm:single-pass max-cut:KK19}
    For every $\epsilon > 0$, every single-pass adversarial-order streaming algorithm which $(\frac12+\epsilon)$-approximates $\MaxCut$ uses $\Omega(n)$ space.
\end{theorem}

\begin{theorem}[\textcite{KKS15}]\label{thm:single-pass max-cut:KKS15}
    For every $\epsilon > 0$, every single-pass random-order streaming algorithm which
    $(\frac12+\epsilon)$-approximates $\MaxCut$ uses $\Omega(\sqrt n)$ space.
\end{theorem}
Note the comparative weaknesses of the two bounds: The first holds only for adversarial-order streams (but in $o(n)$ space), and the second holds only in $o(\sqrt n)$ space (but in randomly-ordered streams).
It is natural to ask whether the limitations in \Cref{thm:single-pass max-cut:KK19,thm:single-pass max-cut:KKS15} are artificial,
or whether we can generalize both bounds simultaneously into a single lower bound:

\begin{conjecture}\label{conj:single-pass max-cut:n-space random-order}
    For every $\epsilon > 0$, every single-pass random-order streaming algorithm which $(\frac12+\epsilon)$-approximates $\MaxCut$ uses $\Omega(n)$ space.
\end{conjecture}

\begin{remark}
There are interesting technical reasons for why assuming adversarial input ordering and/or $o(\sqrt n)$-space makes it easier to prove streaming lower bounds.
We will not delve deeply into lower bound techniques in this column, but we remark that the reasons are ``real'' for other CSPs:
we know that from~\cite{SSSV23-dicut,SSSV23-random-ordering} that
for the related $\MaxDiCut$ problem,
allowing either random input ordering or $\tilde{O}(\sqrt n)$ space strictly increases the achievable approximation ratio
(vs. what $o(\sqrt n)$-space, adversarial-ordering algorithms can achieve).
\end{remark}

All known lower bounds for approximating CSPs via streaming algorithms, including \Cref{thm:single-pass max-cut:KK19,thm:single-pass max-cut:KKS15}, use the following framework:
Define two distributions $\DYes$ and $\DNo$ over streams of constraints,
show that w.h.p. there is a large gap between the $\MaxCSP$ values of the corresponding instances,
and then show that these distributions are indistinguishable in the streaming model of interest via a reduction from a hard one-way communication problem.
Naturally, technical details of the ``source'' communication problem have significant impacts on the exact type of hardness we get for the ``target'' streaming problem (CSP approximation).

In the proof of \Cref{thm:single-pass max-cut:KK19}, $\DYes$ and $\DNo$ have order-sensitive definitions.
More precisely, each stream in the support of $\DYes$ and $\DNo$ can be divided into $O(1)$ successive chunks such that within each chunk, the corresponding edges form a matching.
The input distributions used in \Cref{thm:single-pass max-cut:KKS15} do not have this structure, which turns out to make proving lower bounds hairier.
Morally, this is why the authors of \cite{KKS15} had to ``settle'' for a $o(\sqrt n)$-space lower bound.
This gap between $\sqrt n$ space and $n$ space is a common theme for several of the conjectures in this column.

\begin{remark}
    \Cref{conj:single-pass max-cut:n-space random-order} would imply lower bounds for $(\frac12+\epsilon)$-approximating $\MaxDiCut$
    with single-pass $o(n)$-space random-order streaming algorithms (via the trivial reduction that randomly directs each edge);
    this would demonstrate the tightness of the random-ordering streaming algorithm for $\MaxDiCut$ in \cite{SSSV25}.    
\end{remark}

Another conjecture about lower bounds for $\MaxCut$ with single-pass algorithms is the following:
\begin{conjecture}\label{conj:single-pass max-cut:n log n-space}
    For every $\epsilon > 0$, every single-pass adversarial-ordering streaming algorithm which $(\frac12+\epsilon)$-approximates $\MaxCut$ uses $\Omega(n \log n)$ space.
\end{conjecture}
I.e., we hope to improve over \Cref{thm:single-pass max-cut:KK19} by an additional logarithmic factor in the space usage.
This would match the space usage of the generic sparsifier-based $(1-\epsilon)$-approximation for all CSPs (\Cref{thm:sparsifier}).

\section{Multi-pass streaming lower bounds}\label{sec:multi-pass max-cut}

For a long time, despite some works~\cite{AKSY20,AN21} making partial progress, we seemed very far from any full understanding of the hardness of approximating $\MaxCut$
once algorithms are allowed more than one pass over the input distribution.
This changed with the recent breakthrough work of \textcite{FMW25}, who proved the following amazing result:

\begin{theorem}[{\textcite{FMW25}}]\label{thm:multi-pass max-cut:FMW25}
For every $\epsilon > 0$, every $k$-pass, $s$-space streaming algorithm which $(\frac12 + \epsilon)$-approximates $\MaxCut$ has $ks = \Omega(\sqrt[3]{n})$.
\end{theorem}
The proof of \cite{FMW25} introduces some very novel ideas to the study of streaming CSP approximations,
including an argument which formalizes some folklore intuition about streaming algorithms for $\MaxCut$:
Optimal algorithms essentially just use their memory space to remember increasingly large connected components in the graph,
and then search for odd-length cycles in these components as they keep seeing additional edges.
Thus, the task of proving lower bounds against arbitrary algorithms morally reduces to proving lower bounds only against these algorithms.\footnote{
    \cite{KK19} also includes a very nice analysis of these ``component-growing'' algorithms in the single-pass setting.
    It would be very interesting if the reduction to component-growing protocols in \cite{FMW25} could be reworked into the single-pass setting,
    giving a simpler proof of the \cite{KK19} result for general algorithms.
}

There are numerous interesting questions following up on \cite{FMW25}.
For instance, it is not clear at all what happens once we allow $\omega(\sqrt[3]{n})$ space and $O(1)$ passes.
For starters, we conjecture the following, which would generalize the $o(n)$-space lower bound for a single pass in \Cref{thm:single-pass max-cut:KK19}:

\begin{conjecture}\label{conj:multi-pass max-cut:two-pass n-space}
    For every $\epsilon > 0$, every two-pass, adversarial-order streaming algorithm which $(\frac12 + \epsilon)$-approximates $\MaxCut$ uses $\Omega(n)$ space.
\end{conjecture}

It is also interesting to consider how crucial the $\sqrt[3]{n}$-space threshold in \Cref{thm:multi-pass max-cut:FMW25} is.
Perhaps one could prove the following:
\begin{conjecture}\label{conj:multi-pass max-cut:sqrt(n)-space-pass}
    For every $\epsilon > 0$, every $k$-pass, $s$-space streaming algorithm which $(\frac12 + \epsilon)$-approximates $\MaxCut$ has $ks = \Omega(\sqrt{n})$.
\end{conjecture}
However, to my current knowledge, there are multiple places where the \cite{FMW25} argument breaks beyond $\sqrt[3]{n}$ space, and so proving \Cref{conj:multi-pass max-cut:sqrt(n)-space-pass} may be very hard.

\begin{remark}
    There is a folklore result which shows that the hard distributions $\DYes$ and $\DNo$ used in \cite{FMW25} to prove \Cref{thm:multi-pass max-cut:FMW25}
    (which are roughly the same instances as those used in \cite{KKS15,KK19} to prove \Cref{thm:single-pass max-cut:KK19,thm:single-pass max-cut:KKS15})
    are actually \emph{distinguishable} in $\tilde{O}(\sqrt n)$ space and $\tilde{O}(1)$ passes.
    Very roughly, in this regime, one can take $O(\sqrt n)$ random walks of length $O(\log n)$ in the input graph
    and find odd-length cycles in $\DNo$ via looking at collisions among the walks' endpoints.
    This is why our \Cref{conj:multi-pass max-cut:sqrt(n)-space-pass} goes only up to the $\sqrt{n}$ threshold.
\end{remark}

Beyond $\sqrt{n}$ space, it is much less clear what should happen.
One reasonably safe conjecture might be the following:

\begin{conjecture}\label{conj:multi-pass max-cut:beating 1/2}
    For every $C > 0$, there exists some $\epsilon > 0$ such that every
    streaming algorithm which $(1-\epsilon)$-approximates $\MaxCut$ uses $\Omega(n^C)$ passes or $\Omega(n)$ space.
\end{conjecture}

See also~\cite[Rmk.~1.5]{STV25} for discussion on semidefinite-programming-based multi-pass algorithms for $\MaxCut$.

\section{More $o(\sqrt n)$-space streaming lower bounds}

It turns out that $\MaxDiCut$ behaves very differently than $\MaxCut$ in the streaming setting:
It admits nontrivial approximations, while $\MaxCut$ does not.
Some intuition for this is the following:

\begin{remark}\label{rmk:dicut vs cut}
    A directed graph $\calG$ is satisfiable for $\MaxDiCut$ iff every vertex has either all outgoing or all incoming edges.
    Thus, it is easy to detect \emph{locally} whether a $\MaxDiCut$ instance is not perfectly satisfiable,
    i.e., by just looking at the neighborhood of every vertex independently.
    $\MaxCut$ does not have such a nice characterization:
    A graph $\calG$ is bipartite (a.k.a., is perfectly satisfiable for $\MaxCut$) iff it contains no \emph{odd cycles},
    and so certifying unsatisfiability for $\MaxCut$ requires finding an odd-length cycle,
    which, in a sparse graph, might have length $\Omega(\log n)$.
    Thus, very roughly, it is possible to ``reason locally'' about $\MaxDiCut$, while $\MaxCut$ requires an algorithm to ``reason globally''.
\end{remark}

Building on~\cite{GVV17}, \textcite{CGV20} proved the following characterization for $\MaxDiCut$:
\begin{theorem}[{\textcite{CGV20}}]\label{thm:sqrt-n:CGV20}
    For every $\epsilon > 0$, there is an $O(\log n)$-space sketching algorithm which $(\frac49-\epsilon)$-approximates $\MaxDiCut$,
    but every streaming algorithm which $(\frac49+\epsilon)$-approximates $\MaxDiCut$ uses $\Omega(\sqrt n)$ space.
\end{theorem}

Here the pesky $o(\sqrt n)$-space threshold pops up again.\footnote{
    The \cite{CGV20} algorithm is based on a quantitative form of the observation in \Cref{rmk:dicut vs cut}.
    It measures a quantity called the \emph{average bias} of a directed graph,
    which detects whether typical vertices have either almost all outgoing or almost all incoming edges.}
\textcite{CGSV24} generalized the \cite{CGV20} result into a \emph{dichotomy theorem} between $\tilde{O}(1)$ and $o(\sqrt n)$-space for \emph{sketching} algorithms for \emph{all} CSPs (!):

\begin{theorem}[{\textcite{CGSV24}}]\label{thm:sqrt-n:CGSV24}
    For every $k \in \N$, predicate family $\Pi \subseteq (\{0,1\}^k)^{\{0,1\}}$, and $\alpha \in [0,1]$, either:
    \begin{enumerate}
        \item For every $\epsilon > 0$, there is a sketching algorithm $(\alpha-\epsilon)$-approximating $\MaxCSP$ in $O(\polylog n)$ space.\label{item:upper bound}
        \item  For every $\epsilon > 0$, every sketching algorithm which $(\alpha+\epsilon)$-approximates $\MaxCSP$ uses $\Omega(\sqrt n)$ space.\label{item:lower bound}
    \end{enumerate}
\end{theorem}

\begin{remark}
    \cite{CGSV24} also describes an algorithm for deciding whether \Cref{item:upper bound} or \Cref{item:lower bound} applies,
    which runs in polynomial space in the relevant parameters.
\end{remark}

Note that the general lower bound (\Cref{item:lower bound} in \Cref{thm:sqrt-n:CGSV24}) only holds against \emph{sketching} algorithms.
The authors of \cite{CGSV24} also provide technical conditions under which lower bounds hold more generally against streaming algorithms.
These conditions recover all previously known $o(\sqrt n)$-space streaming lower bounds (\cite{KKS15,GT19,CGV20}),
but we appear far from knowing whether \Cref{item:lower bound} holds against streaming algorithms for all $\Pi$ and $\alpha$.
Hence, it makes sense to examine some CSPs for which the currently-known sketching lower bounds (\`a la \cite{CGSV24}) are stronger than the currently-known streaming lower bounds.

For instance, using the \cite{CGSV24} characterization, \textcite{BHP+22} proved the following:

\begin{theorem}[{\textcite{BHP+22}}]
    For every $\epsilon > 0$, there is an $O(\log n)$-space sketching algorithm which $(\frac29-\epsilon)$-approximates $\MaxThreeAnd$,
    but every \emph{sketching} algorithm which $(\frac29+\epsilon)$-approximates $\MaxThreeAnd$ uses $\Omega(\sqrt n)$ space.
\end{theorem}

A natural follow-up conjecture (which did appear in \cite{BHP+22}) is the following:

\begin{conjecture}\label{conj:sqrt-n:3and}
    For every $\epsilon > 0$, every single-pass \emph{streaming} algorithms which $(\frac29+\epsilon)$-approximates $\MaxThreeAnd$ uses $\Omega(\sqrt n)$ space.
\end{conjecture}

In fact, \cite{BHP+22} contains a general theorem of this form with an explicit constant for $\MaxkAnd$ for every $k$.

\begin{remark}
\cite{BHP+22} does use the \cite{CGSV24} technical condition to show a streaming lower bound for $(\frac29+\epsilon)$-approximating $\MaxThreeAnd$ when $\epsilon > 0.0141$,
but they also show that the condition \emph{cannot} give a full $\frac29$-approximation lower bound.
\end{remark}

\begin{remark}
    I am aware of unpublished work of Raghuvansh Saxena that shows that $\DYes$ and $\DNo$ distributions constructed from the procedure in \cite{CGSV24} for sketching $\MaxThreeAnd$ are indeed distinguishable via streaming algorithms.
    \cite{BHP+22} shows that this pair of the distribution is the unique ``\cite{CGSV24}-type'' pair giving a $(\frac29+\epsilon)$-approximation sketching lower bound.
\end{remark}

Conversely, refuting \Cref{conj:sqrt-n:3and} by demonstrating a streaming algorithm which strictly outperformed all sketching algorithms would of course also be very interesting.

A related example is the \emph{monarchy} function \[
\kMon(x_1,\ldots,x_k) \coloneqq \parens*{ \bigwedge_{i=2}^k x_i } \vee \parens*{ x_1 \wedge \parens*{ \bigvee_{i=2}^k x_i }}.\footnote{
    Think of this as a voting scheme where there is $1$ monarch and $k-1$ subjects.
    $x_1$ is the monarchs's vote and each $x_i$ for $i \in \{2,\ldots,k\}$ is subject $i$'s vote.
    The vote passes if all the subjects vote affirmatively, or if the monarch votes affirmatively and at least one subject does too.
} \]
We define the CSP $\MaxkMon \coloneqq \MaxCSP$ with $\Pi \coloneqq \{\kMon \circ \Not_b\}_{b \in \{0,1\}^k}$.
Note that $\atriv$ for this CSP is $\frac12$.
\textcite{CGS+22-monarchy} studied this (and related) functions vis-a-vis the \cite{CGSV24} dichotomy theorem, and showed the following approximation resistance result:

\begin{theorem}[{\textcite{CGS+22-monarchy}}]
    For every $k \ge 5$ and $\epsilon > 0$, every single-pass \emph{sketching} algorithm which $(\frac12+\epsilon)$-approximates $\MaxkMon$ uses $\Omega(\sqrt n)$ space.
\end{theorem}

We naturally conjecture an analogue of \Cref{conj:sqrt-n:3and} for this problem:

\begin{conjecture}
    For every $k \ge 5$ and $\epsilon > 0$, every single-pass \emph{streaming} algorithm which $(\frac12+\epsilon)$-approximates $\MaxkMon$ uses $\Omega(\sqrt n)$ space.
\end{conjecture}

(See also~\cite[Rmk.~1.7]{STV25} for discussion of algorithms for $\MaxkMon$ which use $o(n)$ space.)
Proving (or refuting) these conjectures would go a long way towards understanding the extent to which the \cite{CGSV24} dichotomy theorem (\Cref{thm:sqrt-n:CGSV24}) characterizes all \emph{streaming} algorithms, not just sketching algorithms.

\section{More lower bounds beyond $o(\sqrt n)$ space}\label{sec:dicut}

There are many further interesting questions that pop up once we move into the regime of $\tilde{O}(\sqrt n)$ and beyond space (but still $o(n)$).
In this regime, we do have strong streaming lower bounds for $\MaxCut$ (\Cref{thm:single-pass max-cut:KK19} due to~\cite{KK19})
and more generally for $\MaxCSP$ when $\Pi$ has the so-called ``wideness'' property~\cite{CGS+22-linear-space}.
At the same time, $\tilde{O}(\sqrt n)$ space is enough to enable some improved approximations,
in particular for $\MaxDiCut$~\cite{SSSV23-random-ordering,SSSV23-dicut} and probably also for $\MaxkAnd$~\cite{Sin23-kand}.
Towards the question of strong $o(n)$-space lower bounds, a very strong conjecture would be the following:
\begin{conjecture}
    Every predicate family $\Pi$ which cannot be nontrivially approximated by $o(\sqrt n)$-space sketching algorithms (\`a la~\cite{CGSV24}) also cannot be nontrivially approximated by $o(n)$-space streaming algorithms.
\end{conjecture}
We would not be surprised if this conjecture is false, but having simple counterexamples would also help orient future study on these types of problems.

Towards the question of $\MaxDiCut$ approximability, \textcite{SSSV23-dicut} 
(building on their earlier work~\cite{SSSV23-random-ordering}, and combined with some numerical work due to \cite{FJ15,Sin23-kand,HSV24}) proved the following theorem:
\begin{theorem}[\textcite{SSSV23-dicut}]\label{thm:beyond sqrt-n:SSSV23}
    There is a $0.485$-approximation single-pass $\tilde{O}(\sqrt n)$-space adversarial-ordering streaming algorithm for $\MaxDiCut$.
\end{theorem}
Note that this is strictly better than the $O(\log n)$-space $(\frac49-\epsilon)$-approximations from \Cref{thm:sqrt-n:CGV20},
which were optimal in $o(\sqrt n)$ space~\cite{CGV20}.

The \cite{SSSV23-dicut} algorithm relies on the notion of ``oblivious algorithms'' from \cite{FJ15}.
For these ``oblivious'' algorithms, lower bounds (at roughly $0.49$) were constructed in \cite{FJ15} and improved in \cite{HSV24}.\footnote{
    The optimal approximation ratio achievable by oblivious algorithms is not currently known;
    the current best word is due to \textcite{HSV24}, who show that the constant is in the interval $[0.4853,0.4889]$.
    Finding (or characterizing) the optimal ratio is an interesting open question.}
It is natural to ask whether one can do better in $o(n)$ space;
\textcite{SSSV25} showed recently that this is possible:

\begin{theorem}[\textcite{SSSV25}]\label{thm:beyond sqrt-n:SSSV25}
    For every $\epsilon > 0$ and $D \in \N$, there is some $\delta > 0$ and a $(\frac12-\epsilon)$-approximation single-pass $\tilde{O}(n^{1-\delta})$-space sketching algorithm for $\MaxDiCut$ on graphs with maximum degree $D$.
\end{theorem}

The maximum degree assumption here is a technical condition which, I believe, can be removed, though it might be quite annoying to do so.
But is the tending-towards-linear space dependence in \Cref{thm:beyond sqrt-n:SSSV25} necessary?
That is, could one even achieve $(\frac12+\epsilon)$-approximations for all $\epsilon > 0$ in $\tilde{O}(\sqrt n)$ space?
We conjecture that this is not possible:

\begin{conjecture}
    There exist constants $\epsilon, \delta > 0$ such that every single-pass sketching algorithm which $(\frac12-\epsilon)$-approximates $\MaxDiCut$ uses $\Omega(n^{\frac12+\delta})$.
\end{conjecture}

If this is true, one could even imagine a whole hierarchy of upper and lower bounds as the approximation ratio tends to $\frac12$ and the space usage tends to $\Theta(n)$:
That is, perhaps, for every $\delta > 0$, there exists $\epsilon > 0$ such that $(\frac12-\epsilon)$-approximating $\MaxDiCut$ is hard in $o(n^{1-\delta})$ space.

\printbibliography

\end{document}